\documentclass[aps,prl,preprint,groupedaddress]{revtex4}
\usepackage{graphicx}
\newcommand{\be}{\begin{equation}}
\newcommand{\ee}{\end{equation}}

\def\AJ{Astrophys. J.}

\def\JGR{J. Geophys. Res.}

\def\PRL{Phys. Rev. Lett.}

\def\SSR{Space Sci. Rev.}

\begin{document}

\title{More than mass proportional heating of heavy ions by supercritical collisionless shocks in the solar corona}

\author{Gaetano Zimbardo}
\affiliation{Physics Department, University of Calabria, Arcavacata di Rende, Italy}

\date{\today}

\begin{abstract}
We propose a new model for explaining the observations 
of more than mass proportional heating of heavy ions in the polar solar corona. 
We point out that a large number of small scale intermittent shock waves can be present
in the solar corona. The energization mechanism
is, essentially, the ion reflection off supercritical quasi-perpendicular collisionless shocks
in the corona and the subsequent acceleration by the motional electric field
${\bf E} = - (1/c) {\bf V} \times {\bf B}$. The acceleration due to ${\bf E}$ is 
perpendicular to the magnetic field, in agreement with observations, and is
more than mass proportional with respect to protons, because the heavy ion orbit
is mostly upstream of the quasi-perpendicular shock foot. The observed temperature
ratios between O$^{5+}$ ions and protons in the polar corona, 
and between $\alpha$ particles and protons in the solar wind are easily recovered.
\end{abstract}

\pacs{}

\maketitle

The heating of the solar corona to temperatures of the order of $10^6$ K and more
is one of the outstanding problems of solar physics. Beside the high temperatures, 
Soho/UVCS observations have shown that heavy ions in polar corona, 
like O$^{5+}$ and Mg$^{9+}$, are heated more than protons,
and that heavy ion heating is more than mass proportional; further, the 
perpendicular temperatures $T_{\perp}$ are much larger than parallel temperatures $T_{\parallel}$
\cite{Kohl97,Kohl98,Cranmer99}.
As a consequence of magnetic mirroring in the diverging magnetic field of 
coronal holes, heavy ions are observed to be faster than protons in the
solar wind \cite{Kohl98,Marsch99}. In addition, the collisional
coupling with protons up to 1.32 $R_{\odot}$ indicates that the 
Mg$^{9+}$ heating has to be faster than minutes \cite{Esser99}. 
These observations give stringent contraints on the coronal heating mechanism.
Ion cyclotron heating has been considered since long (e.g., \cite{Marsch82,Isenberg83,Cranmer99,Hollweg02}), 
but some details are not yet fully understood. 
The comprehension of coronal heating
is of general physical interest, as the sun serves both as a huge plasma laboratory
and as a model for a large class of stars. 

Shock waves are considered to be common in the chromosphere/transition region and in the corona 
(e.g., \cite{Yokoyama95,Yokoyama96,Lee00,Ryutova03,Vecchio09}).
For instance, photospheric convection leads to the emergence of small magnetic loops, which
lead to magnetic reconnection with the network magnetic field; 
small scale plasma jets are formed in the reconnection regions, and fast shocks can form when jets encounter the ambient plasma \cite{Yokoyama96,Tsuneta98,Lee00}. 
Indeed, recent X-ray Hinode and UV Stereo observations have shown that many more plasma jets
are present in the polar corona than previously thought \cite{Cirtain07,Patsourakos09,Nistico09}. 
Therefore, a large number
of small scale, intermittent shocks can form in the reconnection regions and propagate toward the high altitude corona.
Recent numerical simulations show that bursty, time dependent reconnection in solar flares
can eject many plasmoids and create oblique shocks \cite{Tanuma07}. 
In the high corona, magnetic reconnection happens when current
sheets form because of the evolving coronal structures \cite{Fisk03}, while large
scale shocks propagate in the corona because solar flares and of the emergence of coronal mass ejections  \cite{Aschwanden05,Ballai05}. Such shocks are detected as type II radio bursts \cite{Nelson85}.
In the case of solar flares, the associated reconnection outflow termination
shocks can be so strong as to accelerate electrons to 100 keV energies in a fraction of a second
\cite{Tsuneta98,Warmuth09}, and in some cases both the upper and the lower termination shocks
are identified in radio observations \cite{Aurass04}.
 
In the low $\beta$, nearly collisionless corona, a shock wave is formed when a superAlfv\'enic plasma flow having velocity $V_1 > V_A$ collides with the ambient coronal plasma. 
Here, the plasma $\beta$ is given by $\beta = 8\pi p /B^2$,
where $p$ is the total plasma pressure, $B$ is the magnetic field magnitude,   
$V_1$ is the plasma velocity upstream of the shock, and  $V_A = B/\sqrt{ 4 \pi \rho}$
is the Alfv\'en velocity, with $\rho$ the mass density. The Alfv\'enic Mach number is defined as
$M_A = V_1 / V_A$. 
We notice that although the typical Alfv\'en velocity in the
corona, of the order of 1000 km/s, is larger than the observed jet velocities
of 200--800 km/s \cite{Cirtain07,Moreno08}, the Alfv\'en velocity in the reconnection
region can be much lower, since $B$ is weaker there. Indeed, the reconnection
regions are characterized by current sheets, magnetic field reversals,
and magnetic quasi-neutral lines. For instance, Tsuneta and Naito \cite{Tsuneta98} argue that an 
oblique fast shock is naturally formed below the reconnection site in the corona, see their
Figure 1. The plasma velocity in the reconnection outflow region between the slow shocks 
is of the order of $V_A$ in the inflow region, that is much larger than $V_A$ in the outflow region, and this leads to the formation of shocks (e.g., \cite{Yokoyama95,Tsuneta98,Aschwanden05}). 

Previously, shock heating of coronal heavy ions was considered by Lee and Wu \cite{Lee00},
but mostly in connection with subcritical shocks. 
Here, we propose that the more than mass proportional heating of heavy
ions in polar coronal holes is due to ion reflection at supercritical quasi-perpendicular
shocks and to the ion acceleration by the ${\bf V} \times {\bf B}$ electric field in the
shock frame. In this connection, we notice that more than mass proportional heating
of $\alpha$ particles and O$^{6+}$ has been observed in the solar wind by the Ulysses spacecraft, 
between 2.7 and 5.1 AU, downstream of interplanetary shocks, most of which were
supercritical \cite{Berdichevsky97} (see also Ref. \cite{Korreck07}).

It is well known both from laboratory \cite{Paul65,Phillips72} 
and from spacecraft experiments (e.g., \cite{Gosling85,Scudder86,Bale05}) that 
above a critical Mach number $M_A^* \simeq 2.7$  for perpendicular 
collisionless shocks (less than 2.7 if the shock is quasi-perpendicular), a fraction of ions,
which grows with the Alfv\'enic Mach number \cite{Phillips72,Quest86}, 
is reflected off the shock, leading to the
so-called supercritical shocks. When the angle $\theta_{Bn}$ between the shock
normal (pointing in the upstream direction) 
and the upstream magnetic field is larger than about 45$^{\circ}$, the
reflected ions reenter the shock after gyrating in the upstream magnetic field.
Such shocks are termed quasi-perpendicular. Conversely, for $\theta_{Bn} < 45^{\circ}$,
the reflected ions propagare upstream, forming the ion foreshock which characterizes
the quasi-parallel shocks.  The critical Mach number $M_A^*$ can decrease below 1.5
for oblique shocks in a warm plasma \cite{Edmiston84}, so that ion reflection is a
relatively common process. Ion reflection can be considered
to be the main dissipation mechanism by which collisionless shocks convert the flow directed
energy into heat, while the electrons are heated much less (typically, one tenth of proton
heating) \cite{Gosling85}.

\begin{figure}
\includegraphics[width=8.5cm]{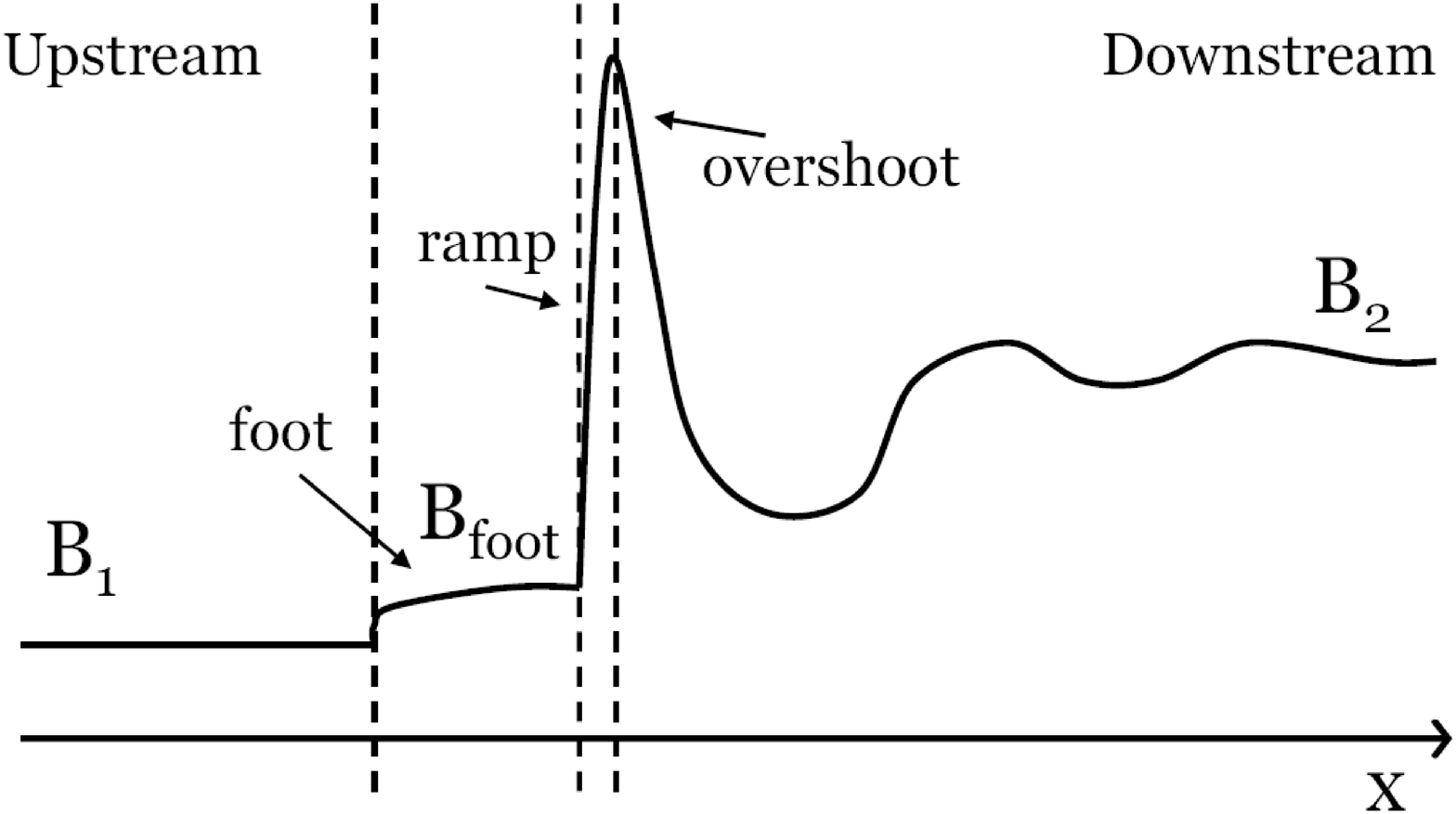}%
\caption{Schematic of the magnetic field profile of a supercritical quasi-perpendicular
collisionless shock. The main features like the magnetic foot, the ramp, and the
magnetic overshoot are indicated. 
\label{f1}}
\end{figure}

For the solar corona, we consider a quasi-perpendicular  
supercritical collisionless shock, and we assume a simple 
one dimensional shock structure. The upstream quantities are indicated
by the subscript 1, and the downstream quantities by the subscript 2.
We adopt the Normal Incidence Frame (NIF) of reference, in which the shock is at rest, 
the upstream plasma velocity is along the $x$ axis and perpendicular to the shock surface, 
${\bf V}_1 = (V_{x1}, 0, 0) $,     
the upstream magnetic field lays in the $xz$ plane, ${\bf B}_1 = (B_{x1}, 0, B_{z1})$,      
so that the motional
electric field $ {\bf E} = - {\bf V} \times {\bf B}/c$ 
is in the $y$ direction, $E_y = V_{x1} B_{z1} /c$.
We further assume that $ B_{z1} \gg B_{x1}$ ($\theta_{Bn} \simeq 90^{\circ}$),
in order to simplify the discussion. 
An order-of-magnitude estimate of the energy gained by ions after reflection
at the shock can be obtained by approximating the reflected ion trajectory 
with a circle of radius $r_L$, with $r_L$ the ion Larmor radius,
in the upstream magnetic field.
Assuming specular reflection \cite{Bale05,Gosling85,Meziane04a},
on average the ion velocity at the reflection point is perpendicular to the shock and along the
$x$ axis. Keeping in mind that ion reflection gives rise to a non adiabatic displacement
in the $y$ direction, the work $W$ done by the electric field is
\be
W = q_i E_y \Delta y,
\ee
where $\Delta y \sim 2 r_L$. For specularly reflected ions,  the Larmor radius has to be evaluated with
the upstream flow speed (neglecting the thermal velocity of the incoming
ion distribution), i.e., $v_{\perp} \simeq |V_{x1}|$, so that
\be
W = q_i E_y \times 2 r_L = 2 q_i  {V_{x1} B_{z1} \over c} {m_i V_{x1} c \over q_i B_{z1} } ,
\ee 
which yields $W = 2 m_i V_{x1}^2$. This estimate shows that the energy gain
is mass proportional.

A more detailed calculation yields a more precise result,
and shows that heavy ion heating is more than mass proportional. 
In order to do this, we remind that a distinctive feature of quasi-perpendicular
collisionless shocks is the formation of a ``foot'' in the magnetic field profile
in front of the main magnetic ramp, the latter culminating in the magnetic overshoot,
beyond which the downstream values are gradually attained \cite{Phillips72,Scudder86,Burlaga08},
see Figure 1. 
The foot is due to the population of
reflected and gyrating protons, which causes an increase in the plasma density,
and, because of the magnetized electrons, in the magnetic field strength \cite{Phillips72}. 
Even if the solar corona composition encompasses several ion species, the foot
extent in the upstream direction is determined by the proton gyroradius, since
protons are the major species. We define $B_{\rm foot} = (1+b) B_{z1}$,
with $b$ depending on the ion reflection rate; $b$ can be estimated to be
of the order 0.5--1 for typical shocks in the heliosphere \cite{Gosling85,Scudder86,Burlaga08}. Direct observations
in space show that very strong fluctuation levels are found in association with
collisionless shocks. However, in what follows we will neglect fluctuations and we will
consider only the average quantities, in order to set the stage.   
Taking into account the fact that the orbit in crossed electric and magnetic fields 
is a trochoid, we start from the equations of the particle trajectory. 
We assume the magnetic field
to be along the $z$ axis, and set the origin of coordinates at the point
of ion reflection, with $t=0$ (e.g., Ref.\ \cite{Landau}):
\be
x(t) = - {v_{\perp}\over \Omega} \sin (\Omega t) + {c E_y \over B} t
\ee
\be
y(t) = {v_{\perp}\over \Omega} [1 - \cos (\Omega t) ]
\ee
where $\Omega = q_i B / m_i c$, and $B$ the local magnetic field. 
The corresponding particle velocity is 
\be
v_x(t) = - {v_{\perp}} \cos (\Omega t) + {c E_y \over B} 
\ee
\be
v_y(t) = {v_{\perp}} \sin (\Omega t) .
\ee
\begin{figure}
\includegraphics[width=8.5cm]{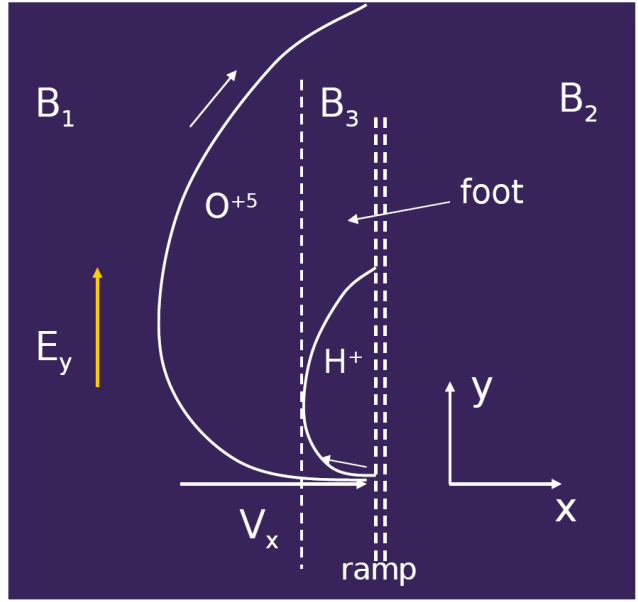}%
\caption{(Color online) Projection in the $xy$ plane of the trajectories of hydrogen and oxygen ions 
reflected at the shock ramp. The
motional electric field is also indicated. As in Figure 1,
the vertical dashed lines separate the main magnetic field regions, such as the upstream
region, the foot, the ramp, and the downstream region.
\label{f2}}
\end{figure}
Specular ion reflection implies that at $t=0$ the ion velocity $v_x$ is
opposite to the incoming plasma velocity, $v_x (t=0) = -V_{x1}$,
whence
\be
v_{\perp} = V_{x1} + {c E_y \over B_{\rm foot}}  \\
          = V_{x1} +  {V_{x1} B_{z1} \over B_{\rm foot}}\\
          = V_{x1} {2+b \over 1+b}  .
\ee
The reflected ions meet again the shock surface, at $x=0$, at a later time $t_1>0$,
corresponding to 
\be
{v_{\perp}\over \Omega} \sin (\Omega t_1) = {c E_y \over B} t_1  .
\ee
Upon inserting the values of $v_{\perp}$ and of $E_y$ in the above equation we obtain
\be
\sin (\Omega t_1) = {\Omega t_1 \over 2+b} ,
\ee
whose numerical inversion yields $\Omega t_1 = 2.27885$ for $b = 1$, 
$\Omega t_1 = 2.1253$ for $b = 0.5$,
and $\Omega t_1 = 1.8955$  for $b=0$ (see below). 
At this time the particle will have moved in the $y$ direction by
an amount given by 
\be
\Delta y(t_1) = {v_{\perp}\over \Omega} [1 - \cos (\Omega t_1) ] =
{m_i V_{x1} c \over q_i B_{z1} } [1 - \cos (\Omega t_1) ]   {2+b \over (1+b)^2}.
\ee
This displacement in the $y$ direction determines the energy gained by reflected ions
during the gyromotion in the field $E_y$:
\be
W = q_i E_y \Delta y = m_i V_{x1}^2  [1 - \cos (\Omega t_1) ] {2+b \over (1+b)^2} 
    =  {2(2+b) \over (1+b)^2} [1 - \cos (\Omega t_1) ] {1\over 2} m_i V_{x1}^2 
\ee
Taking into account the fact that 
 protons move in the foot magnetic field $B_{\rm foot}$, we can assume that
$b\simeq 0.5$--$1$. In such a case,
$1 - \cos (\Omega t_1) = 1.65035$ for $b = 1$ (or 1.52652 for $b=0.5$), so that 
\be
W_{p} \simeq {3\over 2} \times 1.65035 \times ({1\over2} m_p V_{x1}^2) \, .
\ee
On the other hand, for heavy ions like O$^{5+}$ most of the trajectory
is upstream of the foot, see Fig. 2, in the unperturbed plasma where $B \simeq B_{z1}$.
Then we can set $b=0$ with good approximation, and obtain $1 - \cos (\Omega t_1) = 1.319$, so that 
\be
W_{\rm heavy} \simeq 4 \times 1.319 \times ({1\over2} m_i V_{x1}^2) \, .
\ee
Here we can see that, with respect to protons, heating is more than mass proportional.
For $b=1$, the ratio of the heavy ion energy gain over 
the proton energy gain is $W_{\rm heavy} / W_{p} \simeq 2.13 \times m_i/m_p$, while for 
$b=0.5$  we find $W_{\rm heavy} / W_{p} \simeq 1.55 \times m_i/m_p$. Varying the value of
$b$ between 0.5 and 1 yields an O$^{5+}$ temperature about 25--34 times larger than the
proton temperature ($m_{\rm O} \simeq 16 m_p$), 
in good agreement with Soho/UVCS observations which give 
$T_{{\rm O}^{5+}}/T_p = 27$--$37$ \cite{Esser99}. Also, assuming a typical value of 
$b=0.5$, we can easily
recover the temperature ratios observed in the solar wind for helium ($m_{\alpha} \simeq 4 m_p$),
where  $T_{\alpha}/T_p \simeq 6$ is observed in those solar wind periods which are found
to be the least collisional \cite{Kasper08}, thus reflecting more closely the coronal conditions.  
At the same time, for $b=1$, $T_{\alpha}/T_p \simeq 8$ is obtained, a value which is also
observed and reported in Figure 1 of Ref. \cite{Kasper08}. 

On the other hand, 
heating is essentially perpendicular, since it is due to the motional electric field $E_y$ 
which is perpendicular to the magnetic field by definition. This allows to understand
the observed strong temperature anisotropy with $T_{\perp} \gg T_{\parallel}$. 
In addition, a single
shock encounter is required to accelerate the ions, and the acceleration time is on
the scale of the ion gyroperiod, so that the heating mechanism
is very fast, as required by the observations in Ref. \cite{Esser99}.

Typically, the reflection rate for protons is found from numerical simulations
to be 20--30\%. The energy gained by reflected ions is distributed to the transmitted
ions by wave particle interactions \cite{Gosling85}, so that we can assume for the
bulk of protons a heating rate about 1/4 of $W_p$. 
Let us define the heating efficiency, $\eta$, for protons, as the ratio of the
energy gain of both reflected and trasmitted ions over the upstream 
thermal energy ${1\over2} m_p v_{th1}^2$:
\be
\eta = {W_p / 4 \over {1\over2} m_p v_{th1}^2} \simeq 
{2.48 ({1\over2} m_p V_{x1}^2) \over 4 \times {1\over2} m_p v_{th1}^2}
= 0.62 M_s^2
\ee
where $M_s= V_{x1} / v_{th1}$ is the sonic Mach number. In the low $\beta$ corona, the thermal speed
is much less than the Alfv\'en speed, so that $M_s \gg M_A$. However, 
in the reconnection outflow region the magnetic field is weaker than in the ambient corona,
so that 
for a first estimate we assume that the sonic Mach number is of the
same order of the Alfv\'en Mach number. For instance, assuming that $M_s = 7$, we can see that
the efficiency  for protons equals $\eta\simeq 30$. In other words, the crossing of two or three  
shocks might bring the chromospheric plasma from temperatures
of the order of $10^4$ K to coronal temperatures of the order of $10^6$ K.
As a general trend, we can say that the stronger the magnetic field in the reconnection
inflow region, the larger the plasma velocity and the Mach numbers in the reconnection
outflow region, and the larger the heating efficiency. On the other hand, several shock 
crossings may be required for the high altitude corona to reach the observed temperatures,
and we notice that large scale shocks associated with coronal mass ejections and type II radio
bursts can propagate all the way into polar corona. 

Clearly, the present model has to be further developed, since a wide range of different Mach numbers, plasma $\beta$, shock normal angles $\theta_{Bn}$, fluctuation levels, 
and heating ratios can be envisaged in the corona. In addition, collisional coupling with protons
can decrease the obtained temperature ratios, as may be the case of Mg$^{9+}$. 
Further, multiple reflected ions at a single shock can also be envisaged \cite{Lipatov99},
a phenomenon which would increase the heating efficiency.

Our model leads to the prediction that a fraction of heavy ions comparable to the proton
fraction is also reflected at quasi-perpendicular shocks. 
In collisionless shocks, 
an electrostatic potential barrier $\Delta \phi \simeq m_p (V_{x1}^2 - V_{2}^2)/2e$
arises, which slows down the incoming ions \cite{Phillips72,Leroy82,Bale07}. 
In simple discussions of ion reflection,
ions are expected to undergo specular reflection if their kinetic energy in the shock frame 
is less than the potential energy barrier $q_i\Delta \phi$: however, for heavy ions this is usually found only for a tiny fraction of the upstream velocity distribution. 
 Nevertheless, 
experimental evidence of $\alpha$ particle reflection at Earth's quasi-perpendicular bow 
shock has been reported by in Ref. \cite{Scholer81}, while evidence of 
$\alpha$ particle specular reflection off the quasi-parallel bow shock was reported in
Ref. \cite{Fuselier95}. On the other hand,
laboratory experiments show that increasing $M_A$, the number of reflected
ions increases while the potential jump \it decreases \rm \cite{Phillips72}, contrary
to expectations if the potential jump would be the only cause of ion reflection.
This shows that ions are not simply reflected by the average potential jump across the shock, 
and that also the magnetic foot and the fluctuating electric and magnetic overshoots play a role for ion reflection \cite{Leroy82,Meziane04a,Meziane04b,Bale05}. Indeed,  
cross shock electric fields measured by the Polar spacecraft at Earth's bow shock show that 
the potential $\Delta \phi$ is strongly spiky and fluctuating \cite{Bale07}. 
Further, quasi-perpendicular shocks also exhibit cyclic reformation,
which implies time-depending electric and magnetic overshoots and reflection rate 
\cite{Quest86,Lobzin07}. 
Indeed, the Cluster spacecraft have recently shown that ion reflection is highly unsteady \cite{Lobzin07}, so that the strong variations in the shock structure can also induce heavy ion reflection.

In conclusion, we have considered the heavy ion energization
due to the ion reflection off quasi-perpendicular shocks. Fast, supercritical shocks
are assumed to form because of reconnection of small scale magnetic loops at the
base of coronal holes, like those associated with polar coronal jets,
and because of the merging of magnetic structures in the 
higher corona. The energy stored in the coronal magnetic field is transformed
to bulk kinetic energy by reconnection, and into heat and heavy ion heating
by the quasi-perpendicular shocks which form in the reconnection outflow region.  
Our model can explain both
coronal heating and the more than mass proportional 
heavy ion heating observed by Soho/UVCS. In addition, this heating mechanism is strictly perpendicular to the magnetic field and it is very fast (a single shock encounter is needed);
most heating goes into the ions, with electrons undergoing an energy gain
which is about an order of magnitude smaller than that of protons \cite{Gosling85,Scudder86}. 
Further, the strongly anisotropic heating with $T_{\perp} \gg T_{\parallel}$ 
can give rise to efficient ion cyclotron emission; this phenomenon is actually observed downstream of the Earth's bow shock. Indeed, recent Stereo, Helios, and Venus Express data show that ion
cyclotron waves are probably present in the corona \cite{Jain09}. 
Therefore, quasi-perpendicular collisionless shocks can be the source of ion cyclotron
waves in the corona, too. These waves 
 later on can heat locally the solar wind by cyclotron resonance dissipation, as suggested
by a number of observations.

\begin{acknowledgments}
This research was partially supported by the Italian INAF and 
the Italian Space Agency, contract ASI n. I/015/07/0 ``Esplorazione del Sistema Solare".
\end{acknowledgments}


\end{document}